\pdfoutput=1

\documentclass[11pt]{article}

\usepackage[]{acl}

\usepackage{times}
\usepackage{latexsym}

\usepackage[T1]{fontenc}

\usepackage[utf8]{inputenc}

\usepackage{microtype}

\usepackage[utf8]{inputenc}
\usepackage{graphicx}
\usepackage{amsmath}
\usepackage{booktabs}
\usepackage{todonotes}
\usepackage{amssymb}

\usepackage{multirow}
\usepackage{multicol}
\usepackage{microtype}
\usepackage{subcaption}
\usepackage{enumitem,kantlipsum}
\usepackage[normalem]{ulem}
\usepackage{tabularx}
\usepackage{tcolorbox}
\usepackage{color, colortbl}
\usepackage{pbox}
\usepackage{listings}
\usepackage{xstring}
\usepackage{graphicx}
\usepackage{pbox}
\usepackage{subcaption}
\usepackage{epstopdf}
\usepackage{booktabs}
\usepackage{multirow}

\usepackage{hyperref}

\usepackage{hyperref}

\title{A Survey on Multimodal Disinformation Detection}

\author{
Firoj Alam$^1$, Stefano Cresci$^2$, Tanmoy Chakraborty$^{3}$\thanks{\hspace{1.5mm}Work done while T. Chakraborty was at IIIT-Delhi, India.} \hspace{2mm}, Fabrizio Silvestri$^4$,\\ 
\textbf{Dimitar Dimitrov$^5$, Giovanni Da San Martino$^6$, Shaden Shaar$^1$,} \\
\textbf{Hamed Firooz$^7$, Preslav Nakov$^8$} \\
{\small \textbf{$^1$Qatar Computing Research Institute, HBKU, $^2$IIT-CNR, Pisa, $^3$IIT Delhi, India}} \\
{\small \textbf{$^4$Sapienza University of Rome, Italy, $^5$Sofia University, 
$^6$University of Padova, Italy, 
$^7$Facebook AI,}}\\
{\small \textbf{$^8$Mohamed bin Zayed University of Artificial Intelligence}}\\
{\small \textit{\{fialam, sshaar\}@hbku.edu.qa, s.cresci@iit.cnr.it, tanchak@ee.iitd.ac.in, fsilvestri@diag.uniroma1.it}}\\
{\small \textit{mitko.bg.ss@gmail.com, dasan@math.unipd.it,mhfirooz@fb.com, preslav.nakov@mbzuai.ac.ae}}
}

\begin{document}
\maketitle
\begin{abstract}
Recent years have witnessed the proliferation of offensive content online such as fake news, propaganda, misinformation, and disinformation. While initially this was mostly about textual content, over time images and videos gained popularity, as they are much easier to consume, attract more attention, and spread further than text. As a result, researchers started leveraging different modalities and combinations thereof to tackle online multimodal offensive content. In this study, we offer a survey on the state-of-the-art on {\em multimodal disinformation detection} covering various combinations of modalities: text, images, speech, video, social media network structure, and temporal information. Moreover, while some studies focused on {\em factuality}, others investigated how {\em harmful} the content is. While these two components in the definition of disinformation -- (\emph{i})~factuality, and (\emph{ii})~harmfulness --, are equally important, they are typically studied in isolation. Thus, we argue for the need to tackle disinformation detection by taking into account multiple modalities as well as both factuality and harmfulness, in the same framework. Finally, we discuss current challenges and future research directions.
\end{abstract}

\section{Introduction}
The proliferation of online social media has encouraged individuals to freely express their opinions and emotions. On one hand, the freedom of speech has led to a massive growth of online content which, if systematically mined, can be used for citizen journalism, public awareness, political campaigning, etc. On the other hand, its misuse has given rise to the proliferation of hostility online \cite{brooke-2019-condescending,joksimovic-etal-2019-automated}, resulting in offensive content in the form of fake news, hate speech \cite{schmidt-wiegand-2017-survey,davidson2017automated}, propaganda \cite{EMNLP19DaSanMartino}, cyberbullying \cite{van-hee-etal-2015-detection}, etc. 
Indeed, researchers have argued that this situation has set the dawn of the Post-Truth Era, dominated by emotions and ``alternative facts''~\cite{LEWANDOWSKY2017353,cooke2018fake,nakov-da-san-martino-2020-fact}. More recently, with the emergence of the \mbox{COVID-19} pandemic, a new blending of medical and political false information has given rise to the first global infodemic~\cite{paka2021crosssean,zarocostas2020fight,10.1007/978-3-030-73696-5_3}.\footnote{\url{https://www.who.int/health-topics/infodemic}} 

The term ``fake news'' is commonly used, although it is very generic, and misleads people to focus only on veracity. That is why international organizations such as the UN, WHO, EU, and NATO prefer the term {\em \textbf{disinformation}}~\cite{ireton2018journalism}, which refers to information that is (\emph{i})~{\em fake} and also (\emph{ii})~spreads {\em deliberately to deceive and harm} others. The latter aspect of the disinformation (i.e.,~harmfulness) is often ignored, but it is equally important. 
A related term is \emph{misinformation}, which also refers to the spreading of false content, but lacks the underlying intention to do harm. This is illustrated by the definitions of these notions by First Draft
\cite{ireton2018journalism} where \emph{\textbf{misinformation}} is defined as ``\emph{unintentional mistakes such as inaccurate photo captions, dates, statistics, translations, or when satire is taken seriously}'', while \emph{\textbf{disinformation}} is ``\emph{fabricated or deliberately manipulated text/speech/visual context, and also intentionally created conspiracy theories or rumors}''.

In our survey, we will focus on disinformation, and we will study both the factuality and harmfulness aspects of the problem, with focus on different modalities. Note that there are posts that can be harmful but factually true or non-factual but harmful (e.g., hate speech); our study also covers some related work on them. 
The term \emph{factuality} refers to automatically evaluating the solidity of the reporting/social media statements in terms of facts and figures~\cite{ireton2018journalism}. 
 The \emph{harmfulness} or \emph{harmful content}  typically refers to ``{\em anything online which causes a person distress or harm}''.\footnote{\url{https://swgfl.org.uk/services/report-harmful-content/}} Figure~\ref{fig:example_fact_harmful}, in Appendix, gives examples of such content. 
\citet{alam2020fighting} addressed both aspects of disinformation using social media content related to the COVID-19 infodemic. They demonstrated a correlation between factuality and harmfulness, which varies across languages even in the same country, e.g., for Arabic, 56\% of the false content was harmful, while for English, it was 24\%.


Disinformation often spreads as text. However, Internet and social media allow the use of different modalities, which can make a disinformation message attractive as well as impactful, e.g.,~a meme or a video is much easier to consume, attracts much more attention, is perceived as more credible \cite{Hameleers2020}, spreads further than simple text~\cite{zannettou2018origins}, and can be weaponized \cite{olsen2018memes}.

Notably, multimodality remains under-explored in disinformation detection. \citet{bozarth2020toward} performed a meta-review of 23 fake news models and the data modality they leveraged, and found that 91.3\% used text, 47.8\% looked into social media network structure, 26\% relied on temporal data, and only a handful made use of images or videos. Moreover, while there has been research 
to detect whether an image or a video has been manipulated, the attempt is less in a truly multimodal setting~\cite{Perez-Rosas:2015:DDU:2818346.2820758,tan-etal-2020-detecting,zhang2022scenefnd,song2021multimodal,giachanou2020multimodal2,denaux2020linked}.

Here we survey research on multimodal disinformation detection covering various combinations of modalities: text, images, speech, video, social media network structure, and temporal information. The data sources include social media (e.g.,~Twitter), 
news, video 
(e.g., courtroom trials), and TV shows. We further argue for the need to cover multiple modalities in the same framework, while taking both factuality and harmfulness into account.

\begin{figure}[htb!]
\includegraphics[width=0.95\columnwidth]{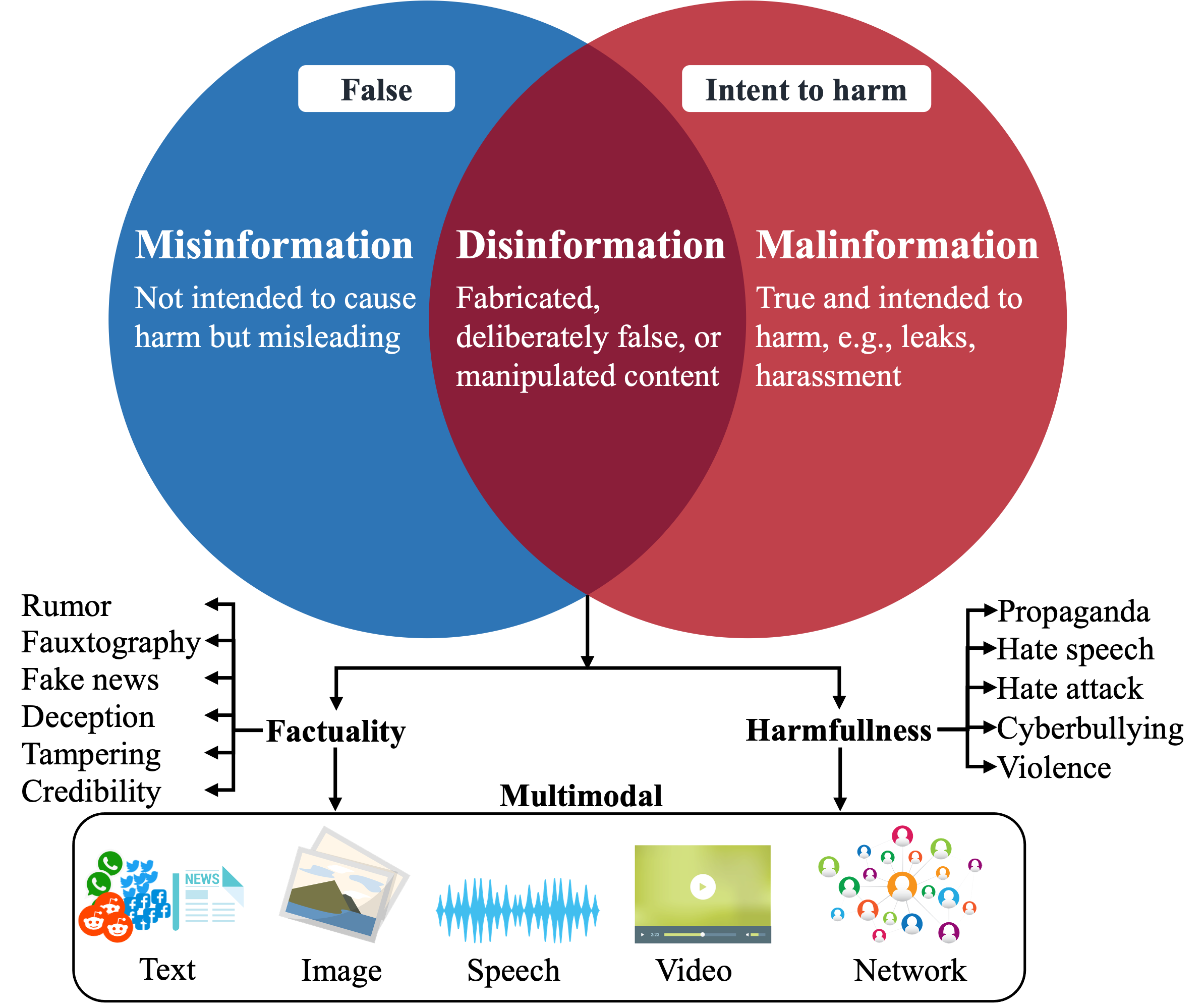}
\centering
\caption{Our vision of multimodality to interact with harmfulness and factuality in this survey.
}
\label{fig:dimensions}
\vspace{-5mm}
\end{figure}

While there have been a number of surveys on ``fake news''~\cite{Shu:2017:FND:3137597.3137600,kumar2018false,CardosoDurierdaSilva2019,zhou2020survey}, misinformation \cite{Islam2020}, fact-checking \cite{thorne-vlachos:2018:C18-1,Kotonya2020}, truth discovery \cite{Li:2016:STD:2897350.2897352}, rumour detection \cite{bondielli2019survey}, harmful memes \cite{sharma2022detecting} and propaganda detection \cite{da2020survey}, none of them had multimodality as the main focus. Moreover, they targeted either factuality (most surveys above), or harmfulness (the latter survey), but not both. Here, we aim to bridge this gap. Therefore, in the present survey, we analyze the literature covering various aspects of multimodality (text, image, speech, video, network, and temporal), with a focus on the two aspects of disinformation: factuality and harmfulness, as shown in Figure~\ref{fig:dimensions}.

\section{Multimodal Factuality Prediction}
\label{sec:factuality}

In this section, we focus on the first aspect of disinformation -- {\em factuality}. Automatic detection of factual claims is important to debunk the spread of misleading information, as it is crucial to detect the factuality of statements that can mislead people. A large body of work has been devoted to fact-checking textual claims but such claims are often expressed and disseminated together with other modalities such as images, speech, and video, and are further propagated through social networks. We summarize relevant studies in Table \ref{tab:related_studies_factuality}. 

\begin{table*}[tbh!]
\centering
\setlength{\tabcolsep}{2.0pt}
\scalebox{0.53}{
\begin{tabular}{@{}llllllllllll@{}}
\toprule
\multicolumn{1}{c}{\textbf{Ref.}} & \multicolumn{1}{c}{\textbf{Task}} & \multicolumn{5}{c}{\textbf{Modality}} & \multicolumn{1}{c}{\textbf{Data/Source}} & \multicolumn{1}{c}{\textbf{Anno.}} & \multicolumn{1}{c}{\textbf{Lang}} & \multicolumn{1}{c}{\textbf{Method}} \\ \midrule
 &  & \multicolumn{1}{c}{\textbf{T}} & \multicolumn{1}{c}{\textbf{I}} & \multicolumn{1}{c}{\textbf{V}} & \multicolumn{1}{c}{\textbf{N}} & \multicolumn{1}{c}{\textbf{S}} &  &  &  &   \\ \midrule
\cite{baly2020written} & Bias, factuality & \checkmark &  &  &  & \checkmark & MBFC & M & En & SVM, BERT \\
\cite{dinkov2019predicting} & \begin{tabular}[c]{@{}l@{}}Bias\end{tabular} & \checkmark &  &  &  & \checkmark & MBFC  & M & En &  MM deep learning architecture \\
\cite{d18-1389} & Bias, factuality & \checkmark &  &  &  &  & MBFC & M & En & SVM \\
\cite{shao2018spread} & Credibility$*$ & \checkmark &  &  &  &  & Articles and tweets & M & En &  Statistical analysis \\
\cite{sen2020multimodal} & Deception &  &  & \checkmark &  &  & CTD: 121 videos & M & En &  RF, SVM and NN classifiers \\
\cite{soldner2019box} & Deception &  &  & \checkmark &  &  & \begin{tabular}[c]{@{}l@{}}TV Show\end{tabular} & M & En &  RF \\
\cite{Volkova_Ayton_Arendt_Huang_Hutchinson_2019} & \begin{tabular}[c]{@{}l@{}}Deception\end{tabular} & \checkmark & \checkmark &  &  &  & \begin{tabular}[c]{@{}l@{}}Twitter; T1: 2,485, \\T2-T3: 56,691,\\ T4: 496,929\end{tabular} & M & En &  Feature fusion with AdaBoost/NN \\
\cite{krishnamurthy2018deep} & Deception &  &  & \checkmark &  &  & CTD: 121 videos & M & En &  MLP \\
\cite{kaya2016fusing} & Deception &  &  &  &  & \checkmark & CSC: 25 videos & M & En &  PLS/ELM based model \\
\cite{levitan2016combining} & Deception &  &  &  &  & \checkmark & CSC: 25 videos & M & En &  SMO, Bagging, Dagging, BN and NB \\
\cite{Perez-Rosas:2015:DDU:2818346.2820758} & Deception &  &  & \checkmark &  &  & CTD: 121 videos & M & En &  DT and RF \\
\cite{hirschberg2005distinguishing} & Deception &  &  &  &  & \checkmark & CSC: 25 videos & M & En &  Rule-based classifier \\

\cite{kazemi-etal-2021-claim} & Facuality & \checkmark &  &  &  &  & FEVER & M & En & Deep Q-learning network \\ 
\cite{atanasova-etal-2020-generating-fact} & Factuality & \checkmark &  &  &  &  & Liar-Plus & M & En & DistilBERT \\
\cite{sathe-etal-2020-automated} & Facuality & \checkmark &  &  &  &  & WikiFactCheck & M & En & SVM, Decomposable attention model \\
\cite{shaar2020known} & Facuality & \checkmark &  &  &  &  & Political debates & M & En &  \begin{tabular}[c]{@{}l@{}}Learning-to-rank approach, BM25, \\ BERT, RoBERTa, sentence-BERT\end{tabular} \\
\cite{kopev2019detecting} & \begin{tabular}[c]{@{}l@{}}Facuality\end{tabular} & \checkmark &  & \checkmark &  &  & Political debates  & M & En &  MM fusion: concatenation \\
\cite{vo2018rise} & Facuality & \checkmark &  &  &  &  & \begin{tabular}[c]{@{}l@{}}Fact-checked tweets from:\\ Snopes.com, Politifact.com, \\ FactCheck.org, OpenSecrets.org, \\ TruthOrfiction.com and \\ Hoax-slayer.net\end{tabular} & M & En & \begin{tabular}[c]{@{}l@{}}BPRMF , MF, CoFactor, CTR, \\ proposed a joint model\end{tabular} \\
\cite{baly-EtAl:2018:N18-2} & Facuality & \checkmark &  &  &  &  & Claims from Verify and Reuters & M & Ar & \begin{tabular}[c]{@{}l@{}}Gradient boosting, multilayer perceptron, \\ softmax layer, end-to-end memory network\end{tabular} \\
\cite{rashkin-EtAl:2017:EMNLP2017} & Facuality & \checkmark &  &  &  &  & Politifact & M & En & LSTM, MaxEnt, NB \\

\cite{10.1145/3340531.3412046} & Fake news & \checkmark &  &  & \checkmark &  & \begin{tabular}[c]{@{}l@{}}PHEME , \\Twitter (snopes.com), Weibo, \\FakeNewsNet\end{tabular} & M & En &  \begin{tabular}[c]{@{}l@{}}Graphical social context\end{tabular} \\ 
\cite{nakamura2019r} & Fake news  & \checkmark & \checkmark &  &  &  & Reddit: 1m posts & DS & En &  MM fusion \\
\cite{shu2020hierarchical} & Fake news & \checkmark &  &  & \checkmark &  & PolitiFact and GossipCop & M & En &  GNB, DT, LR, and RF \\
\cite{shu2019beyond} & Fake news & \checkmark &  &  & \checkmark &  & BuzzFeed and PolitiFact & M & En &  \begin{tabular}[c]{@{}l@{}}LR, NB, DT, \\XGBoost, AdaBoost, and GB\end{tabular} \\
\cite{vosoughi2018spread} & Fake news & \checkmark &  &  & \checkmark &  & Twitter: 126,000 posts & M & En &  Statistical analysis, Topic modeling \\
\cite{liu2018early} & Fake news & \checkmark &  &  & \checkmark &  & \begin{tabular}[c]{@{}l@{}}Weibo: 4,664 \cite{ma2016detecting}, \\ Twitter15: 1,490 \cite{ma2017detect}, \\ Twitter16: 818 \cite{ma2017detect}\end{tabular} & M & En &  DT, SVM, GRU, RF, RNN, CNN \\
\cite{rashkin-EtAl:2017:EMNLP2017} & Fake news & \checkmark &  &  &  &  & \begin{tabular}[c]{@{}l@{}}Gigaword corpus, articles from \\ seven unreliable news sites\end{tabular} & M & En & MaxEnt \\
\cite{boididou2016verifying} & Fake &  & \checkmark & \checkmark &  &  & Social media & M & En & -\\
\cite{gupta2013faking} & Fake news &  &  &  & \checkmark &  & Twitter: 16,117 tweets & M & En &  DT on balanced dataset, NB \\

\cite{wang2020understanding} & Fauxtography & \checkmark & \checkmark &  &  &  & Twitter, 4chan, and Reddit & M & En &  Analytical \\
\cite{zhang2018fauxbuster} & Fauxtography & \checkmark & \checkmark &  &  &  & Reddit: 91, Twitter: 390 & M & En &  Feature fusion with XGBoost \\

\cite{heller2018ps} & Image tampering$**$ &  & \checkmark &  &  &  & Reddit: 102,028 images & A & - & - \\
\cite{garimella2020images} & Misinformation$**$ &  & \checkmark &  &  &  & WhatsApp: 2,500 images & M & - & - \\
\cite{zannettou2018origins} & Memes propagation  & \checkmark & \checkmark &  &  &  & Twitter, Reddit, 4chan, and Gab & DS & - &  Memes analysis \\
\cite{vosoughi2017rumor} & Rumor & \checkmark &  &  & \checkmark &  & Twitter: 113 false and 96 true & M & En &  \begin{tabular}[c]{@{}l@{}}Temporal, propagation \\linguistic, and user \\credibility features\end{tabular} \\
\cite{kwon2017rumor} & Rumor & \checkmark &  &  & \checkmark &  & \begin{tabular}[c]{@{}l@{}}Twitter, snopes.com, and\\ urban-legends.about.com\end{tabular} & M & En &  RF \\
\bottomrule
\end{tabular}
}
\vspace{-2mm}
\caption{Summary of the most relevant works on factuality, covering different modalities and tasks. \textbf{T:Text}, \textbf{I: Image}, \textbf{V:Video}, \textbf{N:Network}, \textbf{S:Speech}. CTD: Courtroom trial dataset, CSC: Columbia/SRI/Colorado Corpus. Anno.: Annotation, M: manual annotation; DS: distant supervision.   
MM: Multimodal, SVM: Support Vector Machine, RF: Random Forest, DT: Decision Tree; NN: Neural Network, MLP: Multi-layer Perceptron, PLS: Partial Least Squares regression; ELM: Extreme Learning Machines, NB: Naïve Bayes, BN: BayesNet, BPRMF: Bayesian Personalized Ranking Matrix Factorization, , MF: Matrix Factorization, CTR: Collaborative Filtering Regression, GNB: Gaussian Naive Bayes; LR: Logistic Regression; GB: Gradient Boosting, GRU: Gated Recurrent Units, RNN: Recurrent Neural Networks, CNN: Convolutional Neural Networks. $*$ Also include botometer features. T1-T4 represents different tasks. $**$ dataset only.}
\label{tab:related_studies_factuality}
\vspace{-5mm}
\end{table*}

\subsection{Text}
\label{ssec:factuality_text}
Due to the availability of large amounts of textual content, research on the text modality is comparatively richer than for other modalities. Notable work in this direction covers fake news spread on social media \cite{Vosoughi1146}, fake news and fact-checking on news media \cite{rashkin-EtAl:2017:EMNLP2017}, fact-checking such as fact-checked URL recommendation model \cite{vo2018rise} to reduce the spread, fact-checking with stance detection \cite{baly-EtAl:2018:N18-2}, factuality of media outlets \cite{baly2020written,d18-1389}, generating justifications for verdicts on claims \cite{atanasova-etal-2020-generating-fact}, and fact-checking claims from Wikipedia \cite{sathe-etal-2020-automated}. There have also been recent efforts for fact-checking from political debates \cite{shaar2020known,claim:retrieval:context:2021,shaar2021assisting,CheckThat:ECIR2022, clef-checkthat:2022:LNCS}, fact-checking with evidence reasoning \cite{si-etal-2021-topic,jiang-etal-2021-exploring-listwise,wan-etal-2021-dqn} and fact-checking by claim matching \cite{kazemi-etal-2021-claim}. Given that there have been surveys on the text modality for fake news/disinformation detection and fact-checking, here we will not go into more detail about the individual studies.

\subsection{Image}
\label{ssec:factuality_image}
Text with visual content (e.g., images) in social media is more prominent as it is more intuitive; thus, it is easier to consume, it spreads faster, it gets 18\% more clicks, 89\% more likes, and 150\% more retweets \cite{zhang2018fauxbuster}. Due to the growing number of claims disseminated with images, in the current literature, there have been various studies that address the visual content with text for predicting misleading information \cite{Volkova_Ayton_Arendt_Huang_Hutchinson_2019}, fake images \cite{gupta2013faking}, images shared with misinformation in political groups \cite{garimella2020images},
and fauxtography \cite{zhang2018fauxbuster,wang2020understanding}. Some of these studies attempt to understand how two different modalities are used. Their analyses show that the extension of text with images increases the effectiveness of misleading content.
\citet{gupta2013faking} highlighted the role of Twitter to spread fake images. This study reports that 86\% tweets spreading fake images are retweets. \citet{garimella2020images} manually annotated a sample of 2,500 images collected from public WhatsApp groups, and labeled them as \textit{misinformation}, \textit{not misinformation}, \textit{misinformation already fact-checked}, and \textit{unclear}; however, experiments were conducted with binary labels: misinformation vs. not-misinformation. 
The authors 
found that violent and graphic images spread faster. \citet{nakamura2019r} developed a multimodal dataset containing 1M posts including text, images, metadata, and comments collected from Reddit. The dataset was labeled with 2, 3, and 6-ways labels. \citet{Volkova_Ayton_Arendt_Huang_Hutchinson_2019} proposed models for detecting misleading information using images and text. 

\textit{Fauxtography} is defined as \textit{``visual images, especially news
photographs, which convey a questionable (or outright false) sense of the events they seem to depict''} \cite{cooper2007concise}. It is also commonly used in social media in different forms such as a fake image with false claims, a true image with false claims. 
\citet{zhang2018fauxbuster} defined that \textit{``a post is a fauxtography if the image of the post (i) directly supports a false claim, or (ii) conveys misinformation of a true claim.''} An example 
is shown in Figure \ref{fig:example_fact_harmful} (in Appendix \ref{sec:app_exm_factuality_harmful}).
\citet{zhang2018fauxbuster} developed FauxBuster to detect fauxtographic social media content, which uses social media comments in addition to the content in the images and the texts. \citet{EMNLP2019:fauxtography} investigated the factuality of claims with respect to images and compared the performance of different feature groups between text and images. 
\citet{wang2020understanding} analyzed fauxtography images in social media posts and found that posts with doctored images increase user engagement in the form of re-shares, likes, and comments, specifically in Twitter and Reddit. They pointed out that doctored images are often used as memes to mislead or as a means of satire, and that they have a `clickbait' power to drive engagement.

\subsection{Speech/Audio}
\label{ssec:factuality_speech_audio}
There have been attempts to use acoustic signals to predict the factuality of claims in political debates \cite{kopev2019detecting,shaar2020known}, left-center-right bias in YouTube channels \cite{dinkov2019predicting}, and deception in speech \cite{hirschberg2005distinguishing}. \citet{kopev2019detecting} found that the acoustic signal helps in improving the performance compared to using only textual and metadata features. Similarly, \citet{dinkov2019predicting} reported that the use of speech signal improves the performance of the system for detecting the political bias (i.e., left, center, right) of Youtube channels.
Moreover, a large body of work was done on deception detection using the acoustic signal. \citet{hirschberg2005distinguishing} created the Columbia-SRI-Colorado (CSC) corpus by eliciting within-speaker deceptive and non-deceptive speech. Their experiments consist of the use of acoustic, prosodic, and a variety of lexical features including 68 LIWC categories, filled pauses, and paralinguistic information (e.g., speaker information, gender, field-pause). 
Using the same corpus, an evaluation campaign was organized, where different multimodal approaches were proposed, such as fusion of different acoustic, prosodic, lexical, and phonotactics representations \cite{levitan2016combining,kaya2016fusing}.

\subsection{Video}
\label{ssec:factuality_speech_video}
In addition to textual, imagery, and speech content, the information in video plays an important role in capturing cues of deceptive behavior. Such cues in videos (e.g., facial expression, gestures) have been investigated in several studies  
for deception detection.
\citet{Perez-Rosas:2015:DDU:2818346.2820758} developed a real-life courtroom trial dataset, which includes 61 deceptive and 60 truthful videos. They explored the use of $n$-gram features from transcripts and non-verbal features (i.e., facial expressions, eyebrows, eyes, mouth openness, mouth lips, and head movements, hand gestures) to classify 
liars and truth-tellers.
\citet{krishnamurthy2018deep} explored textual, speech, and visual features for deception detection. They used a 3D CNN to extract visual features from each frame, spatio-temporal features, and facial expressions such as smile, fear, or stress. 
\citet{soldner2019box} developed a multimodal deception dataset using TV shows and experimented with textual, 
visual 
and dialog features.


\subsection{Network and Temporal Information}
\label{ssec:factuality_social_network}
The rationale for leveraging network information stems from early work \cite{shao2018spread,vosoughi2018spread} that showed that propagation and interaction networks of fake news are deeper and wider than those of real news. \citet{vosoughi2018spread} further found that fake information spreads faster than factual one, thus advocating for the use of temporal information. 


Propagation networks can be homogeneous or heterogeneous (e.g., encompassing news articles, publishers, users, and posts) and they can be analyzed at different scales 
(e.g., node-level, ego-level, triad-level, community-level and the overall network, as shown in Figure \ref{fig:example_social_network}, in Appendix)~\cite{zhou2019network}.
\citet{shu2020hierarchical} tackled the fake news classification task by proposing an approach based on hierarchical propagation networks. At both micro- and macro-scale, they extracted and jointly considered network features, temporal features, and linguistic features. Experiments on PolitiFact and GossipCop datasets revealed that temporal features have maximum contribution, followed by network and linguistic features.
\citet{shu2019beyond} provided one of the most thorough multimodal frameworks for fake news classification. 
Their experimental results suggest that social context (i.e., network-derived) features are more informative than news content ones. 

\citet{vosoughi2017rumor} proposed Rumor Gauge, a system that jointly exploits temporal and propagation features, in conjunction with linguistic and user credibility features, for checking the veracity of rumors. In particular, Rumor Gauge leverages text, 
and network propagation.
The temporal modality does not directly provide features, but is instead considered by recomputing all other features at regular time steps, thus yielding multiple time series. 
Results by \citet{vosoughi2017rumor} and \citet{kwon2017rumor} also demonstrated that the contribution of the different data modalities change over time.

To mitigate the ``cold start'' problem of propagation-based early detection of fake news, \citet{liu2018early} proposed an approach that is primarily based on user and temporal information. First, they built a propagation path of each news as a time series of user representations. The time series for a given news only contains the ordered representations of those users that shared such news. Then, they learned two vector representations of each propagation path via GRUs and CNNs, respectively. 
\citet{zannettou2018origins} analyzed different aspects of memes, such as how they evolve and propagate in different mainstream and fringe web communities, and variants of memes that propagate. 
Finally, \citet{10.1145/3340531.3412046} proposed Factual News Graph (FANG) to exploit the social structure and the engagement patterns of users for fake news detection.

\section{Multimodal Harmful Content Detection}
\label{sec:harmfulness}

In this section, we focus on the second aspect of disinformation: {\em harmfulness}. It is essential to filter or to flag online harmful content. 
The harmful content includes \textit{child abuse material}, \textit{violent and extreme content}, \textit{hate speech}, \textit{graphic content}, \textit{sexual content}, 
and \textit{spam content}   \cite{banko-etal-2020-unified}.\footnote{\url{https://swgfl.org.uk/services/report-harmful-content/}} In recent years, the ability to recognize harmful content within online communities has received a lot of attention by researchers \cite{pramanick-acl,pramanick2021momenta} and policymakers that aim to keep users safe in the digital world. Studies in this direction include detecting harmful contents in network science \cite{ribeiro2018characterizing}, natural language processing \cite{Waseem2017understanding,Schmidt2017survey,Fortuna2018survey} and computer vision \cite{Yang2019exploring,Vijayaraghavan2019multimodalhate,Gomez2020exploring,SemEval2021-6-Dimitrov}. 
In Table \ref{tab:related_studies_harmfulness}, we provide a list of relevant work addressing different types of harmful content, modalities, source of data, annotation approach, language of the content and the methods.

\begin{table*}[!htb]
\centering
\scalebox{0.53}{
\setlength{\tabcolsep}{2.5pt}
\begin{tabular}{@{}lllllllllll@{}}
\toprule
\multicolumn{1}{c}{\textbf{Ref}} & \multicolumn{1}{c}{\textbf{Task}} & \multicolumn{5}{c}{\textbf{Modality}} & \multicolumn{1}{c}{\textbf{Data/Source}} & \multicolumn{1}{c}{\textbf{Anno.}} & \multicolumn{1}{c}{\textbf{Lang}} & \multicolumn{1}{c}{\textbf{Method}} \\ \midrule
 &  & \textbf{T} & \textbf{I} & \textbf{V} & \textbf{N} & \textbf{S} &  &  &  &  \\ \midrule
\cite{nizzoli2020coordinated} & CIB &  &  &  & \checkmark &  & Twitter: 1.1m users, 11m tweets & DS & En & Statistical and similarity analysis \\
\citet{weber2020whos} & CIB &  &  &  & \checkmark &  & Twitter & -- & En & Statistical and network analysis \\
\cite{9206663} & Cyberbullying & \checkmark & \checkmark &  &  &  & Posts: Vine (970), Instagram (2,218) & M & En & SVM, NB, LR, RF, LSTM, CNN \\
\cite{soni2018see} & Cyberbullying & \checkmark & \checkmark &  &  & \checkmark & Vine videos & M & En & KNN, SVM, LR, RF, GNB \\
\cite{dadvar2018cyberbullying} & Cyberbullying & \checkmark &  &  &  &  & Youtube 54k posts & M & En & LSTM, BiLSTM, CNN \\
\citet{Hosseinmardi2015cyberbull} & Cyberbullying & \checkmark & \checkmark &  & \checkmark &  & Instagram & M & En & SVM \\
\cite{rafiq2015careful} & Cyberbullying & \checkmark &  &  & \checkmark &  & Vine videos & M & En & NB, AdaBoost, DT and RF \\
\cite{van-hee-etal-2015-detection} & Cyberbulling & \checkmark & & & & & Ask.fm: 85k QA pairs & M & Nl & SVM \\
\cite{chatzakou2019detecting} & \begin{tabular}[c]{@{}l@{}}Cyberbullying, \\ Cyberaggression\end{tabular} & \checkmark &  &  & \checkmark &  & Twitter: 1,303 users, 9,484 tweets & M & En & \begin{tabular}[c]{@{}l@{}}NB, RF, AdaBoost, \\ Ensemble, NN\end{tabular} \\
\cite{liang2017temporal} & Gunshots &  &  &  &  & \checkmark & \begin{tabular}[c]{@{}l@{}}Videos: freesound.com, Youtube; \\Test: CSV, TRECVID Gunshot, \\UrbanSound Gunshot\end{tabular} & DS & En & Localized self-paced reranking \\
\cite{mariconti2019you} & Hate attacks & \checkmark & \checkmark &  &  & \checkmark & Youtube videos (428) & M & En & Ensemble, CNN, RNN \\
\cite{kiela2020hateful} & Hate speech & \checkmark & \checkmark &  &  &  & FB: Hateful Memes Challenge  & M & En & \begin{tabular}[c]{@{}l@{}}Late fusion, Concat BERT, MMBT, \\ ViLBERT, VisualBERT\end{tabular} \\
\cite{das2020detecting} & Hate speech & \checkmark & \checkmark &  &  &  & FB: Hateful Memes Challenge  & M & En & VisualBERT, MM fusion \\
\cite{Gomez2020exploring} & Hate speech & \checkmark & \checkmark &  &  &  & Twitter: MMHS150K  & M & En & Inception v3, LSTM, and MM fusion \\
\cite{Yang2019exploring} & Hate speech & \checkmark & \checkmark &  &  &  & FB: train+dev 378k, test 53k & M & En & Fusion: text + image embedding \\
\cite{waseem-hovy-2016-hateful} & Hate speech & \checkmark & & & & & Twitter: 16,914 tweets & M & En &  LR \\
\cite{davidson2017automated} & Hate speech & \checkmark & & & & & Twitter: 24,802 tweets & M & En &  LR, SVM, NB, DT, RF \\
\cite{qian-etal-2018-hierarchical} & Hate speech & \checkmark & & & & & Twitter: 40 accounts, 3.5m tweets & DS & En &  LR, SVM, Char-CNN, BiLSTM, HCVAE \\
\cite{ribeiro2018characterizing} & Hate speech & \checkmark & & & \checkmark & & Twitter: 4,972 users & M & En & GradBoost, AdaBoost, GraphSage \\
\cite{mathew2019spread}  & Hate speech & \checkmark & & & \checkmark & & Gab: 21m posts by 341k users & DS & En & Lexicon based filtering, DeGroot \\
\citet{SemEval2021-6-Dimitrov} & Propaganda & \checkmark & \checkmark &  &  &  & \begin{tabular}[c]{@{}l@{}}FB: SemEval-2021 task 6: \\ 950 Facebook memes\end{tabular} & M & En & MM fusion, MM joint representation \\
\cite{Vijayaraghavan2019multimodalhate} & Hate speech & \checkmark &  &  & \checkmark &  & \begin{tabular}[c]{@{}l@{}}In-house developed and\\ curated datasets\end{tabular} & M & En & \begin{tabular}[c]{@{}l@{}}MM late fusion, LR, SVM, \\ CNN, BiGRU, BiLSTM\end{tabular} \\
\cite{9064936} & Violence  &  & \checkmark & \checkmark &  & \checkmark & VSD96: Hollywood, Youtube & M & En & \begin{tabular}[c]{@{}l@{}}MM Early fusion; SVM, HMM, GMM, \\Bayesian, MLP, QDA, PLDA, CNN, \\KNN, unsupervised, hybrid\end{tabular} \\
\cite{10.1145/2502081.2502187} & Violence  &  &  & \checkmark &  & \checkmark & MediaEval VSD  & M & En & SVM (mid-level audio + low-level visual) \\
\cite{giannakopoulos2009study} & Violence  &  &  &  &  & \checkmark & Movies & M & - & BN, kNN \\
\bottomrule
\end{tabular}
}
\caption{
Summary of the most relevant works on harmful content. 
\textbf{T:Text}, \textbf{I: Image}, \textbf{V:Video}, \textbf{N:Network}, \textbf{S:Speech}, Anno.: Annotation, 
CIB: Coordinated Inauthentic Behavior, QA: Question-answer, CSV: Real-life Conflict Scene Videos, VSD: Violent Scene Detection. Nl: Dutch. KNN:  k-Nearest Neighbors, LSTM: Long Short-Term Memory, BiLSTM: Bidirectional LSTM, MMBT: MultiModal BiTransformers, HCVAE: Hierarchical Conditional Variational Autoencoder, QDA: Quadratic Discriminant Analysis, PLDA: Probabilistic Linear Discriminant Analysis.}
\label{tab:related_studies_harmfulness}
\vspace{-4mm}
\end{table*}

\subsection{Text}
In the past few years there has been significant research effort on detecting harmful content (e.g., hate speech) from social media posts \cite{van-hee-etal-2015-detection,waseem-hovy-2016-hateful,Waseem2017understanding,Schmidt2017survey}. 
\citet{waseem-hovy-2016-hateful} developed a dataset of hate speech consisting of 16K
tweets, and reported a baseline results using char- and word- ngrams and a logistic regression classifier.
 \cite{davidson2017automated} distinguished between hate speech, and offensive language. They developed a dataset of $\sim$24K labeled tweets with categories such as hate speech, offensive language and neither. 
\citet{qian-etal-2018-hierarchical} took a different approach to classic hate speech classification. Instead of binary classes, they proposed 13 fine-grained hate categories such as nationalist, anti-immigrant, racist skinhead, among others, providing a dataset of tweets collected from 40 hate groups.
\citet{ribeiro2018characterizing} proposed an approach to find hateful users on Twitter. \citet{mathew2019spread} analyzed 341K users and 21M posts collected from Gab to understand the diffusion dynamics of hateful content. Their findings suggest that the posts from hateful user diffuse faster, wider, 
and have a greater outreach compared to the posts from non-hateful ones. 

\subsection{Image}
Among different types of harmful content, cyberbullying is one of the major growing problems, significantly affecting teens. \citet{Hosseinmardi2015cyberbull} investigated Instagram images and their associated comments for detecting cyberbullying and online harassment. They developed a manually labeled dataset using CrowdFlower (which is now Appen), where they followed standard procedures for the annotation: using annotation guidelines, qualification tests, gold standard evaluation and quality control criteria such as minimum annotation time. The annotated dataset consists of 998 media sessions (images and their associated comments). A key finding of this study is that a large fraction of the annotated posts (48\%) with a high percentage of negative words have not been labeled as cyberbullying. To train and to evaluate the model, the authors used $n$-grams from text, meta-data (e.g., the number of followers, followees, likes, and shared media), and image categories as features and experimented with Na\"{\i}ve Bayes and SVM classifiers. Their study suggests that combining multiple modalities helps to improve the performance of the SVM classifier.

Hate speech is another important problem that spreads over social media. The ``Hateful Memes Challenge'' is an important milestone to advance the research on this topic and the tasks is to detect hateful memes \cite{kiela2020hateful}. 
\citet{das2020detecting} proposed different approaches for hatefulness detection in memes such as (\emph{i})~extract the caption and include this information with the multimodal model, (\emph{ii})~use sentiment as an additional feature with multimodal representations. 
For hate speech detection, \citet{Yang2019exploring} explored different fusion techniques such as concatenation, bilinear, gated summation, and attention, and reported that combining the text with image embedding boosted the performance in all cases.
\citet{Vijayaraghavan2019multimodalhate} proposed methods for interpreting multimodal hate speech detection models, where the modalities consist of text and socio-cultural information rather than images. Concurrently, \citet{Gomez2020exploring} introduced a larger dataset of 150K tweets for multimodal hate speech detection, consisting of six labels.

Propaganda is another topic that has been explored in multimodal settings. \citet{seo2014visual} showed how Twitter was used as a propaganda tool during the 2012 Gaza conflict to build international support for each side of the conflict. \citet{SemEval2021-6-Dimitrov} addressed the detection of persuasion techniques in memes. 
Their analysis of the dataset showed that while propaganda is not always factually false or harmful, most memes are used to damage the reputation of a person or a group of people. \citet{dimitrov2021detecting} highlighted the importance of both modalities for detecting fine-grained propaganda techniques, with VisualBERT yielding 19\% improvement compared to using the image modality only (with ResNet-152), and 11\% improvement compared to using the text modality only (with BERT). Similar observations were made by \cite{kiela2020hateful} for hateful meme detection. \citet{Glenski2019MultilingualMD} explored multilingual multimodal content and categorizes disinformation, propaganda, conspiracy, hoax, and clickbait.

\subsection{Speech/Audio}
Cues in spoken content can represent harmful behaviors and those cues can be used to automatically detect such content. Due to the lack of data, studies using the speech-only modality are comparatively lower than other modalities even though it plays a major role in many contexts. For example, for detecting violent content such as screaming and gunshots, the speech modality can play an important role, which other modalities might not be able to offer. 
This is important as most often user-generated contents are posted on newspapers or their social media accounts without verifying the content of the post, which can have serious consequences~\cite{harkin2012deciphering,doi:10.1177/2056305117717888}.

\citet{giannakopoulos2009study} studied the audio segmentation approaches for segmenting violent (e.g., gunshots, screams) and non-violent (e.g., music, speech) content in movies. The studies related to violent content detection using acoustic features also include \cite{10.1145/2502081.2502187}, where the focus was on finding violent content in movies.

\citet{liang2017temporal} proposed Localized Self-Paced Reranking (LSPaR) for detecting gunshots and explosion in videos using acoustic features.
\citet{soni2018see} investigated audio, visual and textual features for cyberbullying detection. Their findings suggest that audio and visual features are associated with the occurrence of cyberbullying, and both these features complement textual features.

\subsection{Video}
There are multiple studies on detecting cyberbullying  in video-based social networks such as Vine \cite{rafiq2015careful} and YouTube \cite{dadvar2018cyberbullying}. These studies show that although the percentage of cyberbullying in video sessions is quite low, automatic detection of these types of content is very challenging. 
\citet{9206663} used textual, visual, and other meta-information to detect social media posts with bullying topics. Their proposed method was evaluated on publicly available multimodal cyberbullying datasets. 
\citet{abd2016emotion} investigated the relationship between emotion and propaganda techniques in Youtube videos. Their findings suggest that propaganda techniques in Youtube videos affect emotional responses.
Content (e.g., Youtube videos) can also be attacked by hateful users via posting hateful comments through a coordinated effort. \citet{mariconti2019you} investigated whether a video is likely to be attacked using different modalities such as metadata, audio transcripts, and thumbnails. 

There has been a recent interest from different government agencies to stop the spread of violent content. \citet{9064936} developed a multimodal dataset, which consists of more than 96 hours of Hollywood and YouTube videos and high variability of content. Their study suggests that multimodal approaches with audio and images perform better.

\subsection{Network and Temporal Information}
The use of network data for predicting factuality was motivated by results showing different propagation patterns for fake vs. real content. Such results are lacking for harmful content. However, the intention to harm in social media is often pursued via coordinated actions, for instance, by groups of users (e.g., social bots and trolls~\cite{cresci2020decade}) that target certain people or minorities. These collaborative harmful actions, perpetrated to increase the efficacy of the harm, are best addressed using network analysis to detect likely coordinated harmful campaigns.
\citet{chatzakou2019detecting} focused on detecting cyberbullying and cyberaggression by training machine learning models for detecting: (\textit{i})~bullies, (\textit{ii})~aggressors, (\textit{iii})~spammers, and (\textit{iv})~normal users on Twitter. To solve these tasks, they leveraged a combination of 38 features extracted from user profiles, the textual content of their posts, and network information (e.g., user degree and centrality measures in the social graph). Orthogonal and in synergy with respect to the detection of disinformation, scholars have recently focused on the novel task of detecting Coordinated Inauthentic Behavior (CIB)~\cite{nizzoli2020coordinated}. \textit{CIB is defined as coordinated activities that aim to mislead and manipulate others.}\footnote{\url{https://medium.com/1st-draft/how-to-improve-our-analysis-of-coordinated-inauthentic-behavior-a4ec62ce9bff}}
Detecting CIB typically involves analyzing both interaction networks to detect suspicious coordination, as well as the coordinated users and the content they shared to detect inauthentic users and harmful content \cite{nizzoli2020coordinated,nizzoli2020charting,pacheco2021uncovering}.
Given the importance of coordination in CIB, the analysis typically starts from the available network data by applying community detection algorithms, and subsequently moving to the analysis of textual data.

\section{Modeling Techniques}
\label{sec:modeling}

In this section, we discuss modeling techniques for both factuality and  harmfulness.
To combine multiple modalities, there have been several approaches: {(i) \em early-fusion}, where low-level features from different modalities are learned, fused, and fed into a single prediction model \cite{jin2016novel,yang2018ti,zhang2019multi,spotfake,zhou2020safe,kang2020multi};
{(ii) \em late-fusion}, where unimodal decisions are fused with some mechanisms such as averaging and voting \cite{agrawal2017multimodal,qi2019exploiting}, and {(iii) \em hybrid-fusion}, where a subset of learned features are passed to the final classifier (early-fusion), and the remaining modalities are fed to the classifier later (late-fusion) \cite{jin2017multimodal}. Within these fusion strategies, the learning setup can also be divided into {\em unsupervised}, {\em semi-supervised}, {\em supervised} and {\em self-supervised} methods. 

\citet{SemEval2021-6-Dimitrov} investigated different fusion strategies (e.g., \textit{early}- and \textit{late}-fusion and \textit{self-supervised} models) for propaganda detection using VisualBERT \cite{li2019visualbert}, MMBT \cite{kiela2019supervised}, and ViLBERT \cite{lu2019vilbert}. Their findings suggest that self-supervised joint learning models, such as MMBT, ViLBERT, and VisualBERT perform better in increasing order, respectively, compared to the other fusion methods. 
As a part of ``Hateful Memes Challenge'' to classify hateful vs. non-hateful memes, several such models have been investigated by \citet{kiela2020hateful}.

Attempts to design \textit{unsupervised} models are limited. \citet{muller2020multimodal} introduced Cross-modal Consistency Verification Tool (CCVT) to check the coherence between images and associated texts. \citet{yang2019unsupervised} defined trust of news and credibility of users who spread the news and used Bayesian learning to iteratively update these quantities. News with low trustworthiness is returned as fake news. \citet{gangireddy2020unsupervised} proposed GTUT, a graph-based approach that exploits the underlying bipartite network of users and news articles to detect the dense communities of fake news and fraud users. 

Due to the scarcity of labeled data, a few studies attempted to design \textit{semi-supervised} methods by leveraging an ample amount of unlabelled data. \citet{helmstetter2018weakly,gravanis2019behind} presented weak-supervision and 
\citet{guacho2018semi} presented a tensor-decomposition semi-supervised method for fake content detection. \citet{dong2020two} developed a deep semi-supervised model via two-path learning (one path uses a limited labeled data, the other path explores the unlabelled data) for timely fake news detection. \citet{paka2021crosssean} presented, Cross-SEAN, a cross-stitch semi-supervised end-to-end neural attention model for COVID-19 fake news detection. 

Within a \textit{supervised} learning setup, two other types of learning method have also been explored for disinformation detection such as {\em adversarial learning} and {\em autoencoder based}. \textit{Adversarial learning} models for fake news detection include EANN \cite{wang2018eann}, an event adversarial neural network to detect emerging and time-critical fake news, and SAME \cite{cui2019same}, a sentiment-aware multimodal embedding method which 
leverages multiple modalities with the sentiment expressed by readers in their comments. 


\section{Major Challenges}
\label{sec:major_challenges}

Recently, several initiatives were undertaken by major companies and government entities to combat disinformation in social media \cite{codedisinfo2021}.\footnote{For example, \url{http://digi.org.au/disinformation-code/}} However, automatic detection of misleading and harmful content poses a number of challenges as discussed below and in Appendix (Section \ref{sec:app_more_major_challenges}). 

\noindent\textbf{Models Combining Multiple Modalities.} The major challenge is to devise a mechanism to combine multiple modalities in a systematic way so that one modality complements the others. 
Current state-of-the-art primarily adopts early and late fusion, which are limited and do not always yield strong results~\cite{dimitrov2021detecting}. Very recently, jointly trained multimodal transformer-based models (e.g., ViLBERT \cite{lu2019vilbert}, Visual BERT \cite{lin2014microsoft} and Multimodal Bitransformers (MMBT) \cite{kiela2019supervised}) have shown strong potential \cite{SemEval2021-6-Dimitrov,dimitrov2021detecting,kiela2020hateful}. However, such models are trained considering only two modalities (textual and visual), while fact-checking or disinformation-related content consists of more than two modalities e.g., text, speech, video, network, etc. \cite{baly2020written}. Hence, there is a room for improvement in developing multimodal models 
that involve 
additional, and potentially 
more than two modalities. Another important problem is cross-modal inconsistency in social media content,
as shown in Figure~\hyperref[fig:example_fact_harmful]{\ref{fig:example_fact_harmful}(c)}, 
which poses a challenge in a multimodal setting \cite{tan-etal-2020-detecting}.

\noindent\textbf{Datasets.} 
One of the major challenges when working with such diverse modalities, i.e., text, image, speech, video, and network, is to get access to an appropriate dataset, and moreover to one that considers both factuality and harmfulness.
Furthermore, there is a need to integrate data from multiple platforms (e.g., news, posts
from Twitter, 
Reddit and Instagram) 
as different data sources present different styles and focus on different topics.

\section{Future Directions}
\label{sec:future_forcasting}

Based on the aforementioned challenges, we forecast the following research directions:

\noindent\textbf{Explainability.} Model interpretation remains largely unexplored. This can be addressed in future studies to understand the general capability of the models. Providing evidence of why certain claims are false is also important. There has been work in this direction such as TabFact \cite{Chen2020TabFact:} and FEVER \cite{hanselowski2018ukp}. However, such approaches rely on existing knowledge bases (e.g., Wikipedia) and may fail for a new problem such as disinformation about COVID-19. It is also important to understand what models learn, e.g., lexical or semantic concepts or a set of neurons may learn one aspect better than the others. Moreover, while current studies on explainable fact-checking focus on explaining the predictions, very few focus on model explanations \cite{kotonya-toni-2020-explainable}.

\noindent\textbf{Beyond Content and Network Signals.} 
State-of-the-art methods for multimodal factuality prediction and harmful content detection are primarily based on content signals and network structure. However, the information in these signals is limited and does not include personal preferences or cultural aspects. In the future, we envision multimodal techniques for disinformation detection that would go beyond content and network and would include signals like common sense and information about the user. Moreover, multimodal models will become larger with more heterogeneous signals as input, and they would be pre-trained on a wider variety of tasks to shelter both aspects of disinformation: factuality and harmfulness.

\noindent\textbf{Knowledge-based Method.} The use of knowledge-based approaches to check the factuality of claims based on what has been checked before could be ideal solutions as some claims are often repeated by politicians. 
Current approaches in this direction are limited and this can be explored further by creating a common repository of previously fact-checked claims and harmful content. Relevant studies 
in this direction 
include detecting previously fact-checked claims \cite{shaar2020known}, studying the role of context 
at the sentence level \cite{claim:retrieval:context:2021} or at the document level \cite{shaar2021assisting}, and 
claim matching 
across languages  
\cite{kazemi-etal-2021-claim}. 

\section{Conclusion}
\label{sec:conclusion}
We surveyed the state-of-the-art in multimodal disinformation detection based on prior work on different modalities, focusing on disinformation, i.e.,~information that is both false and intents to do harm. We covered the major research topics of factuality and disinformation. Our survey brought several interesting research challenges for multimodal disinformation detection, such as combining various modalities, which are often not aligned and are in different representations (e.g., text vs. speech vs. network structure), and the lack of such datasets to foster future research. In addition to highlighting the challenges, we also pointed to several research directions. While doing so, we  argued for the need to tackle disinformation detection by taking into account multiple modalities as well as both factuality and harmfulness in the same framework.

\section*{Acknowledgments}
This research is part of the Tanbih mega-project, developed at the Qatar Computing Research Institute, HBKU, which aims to limit the impact of ``fake news,'' propaganda, and media bias by making users aware of what they are reading, thus promoting media literacy and critical thinking. T Chakraborty would like to thank the support of Wipro Research Grant. 

\section*{Limitations}
We might not have covered \emph{all} relevant work that falls under the topics of factuality and harmfulness across different modalities.

\bibliographystyle{acl_natbib}
\bibliography{bib/all_bib}

\section*{Appendix}
\label{sec:appendix}
\appendix

\section{Examples of Factuality and Harmful Content}
\label{sec:app_exm_factuality_harmful}

\begin{figure}[h]
\centering
\includegraphics[width=0.42\textwidth]{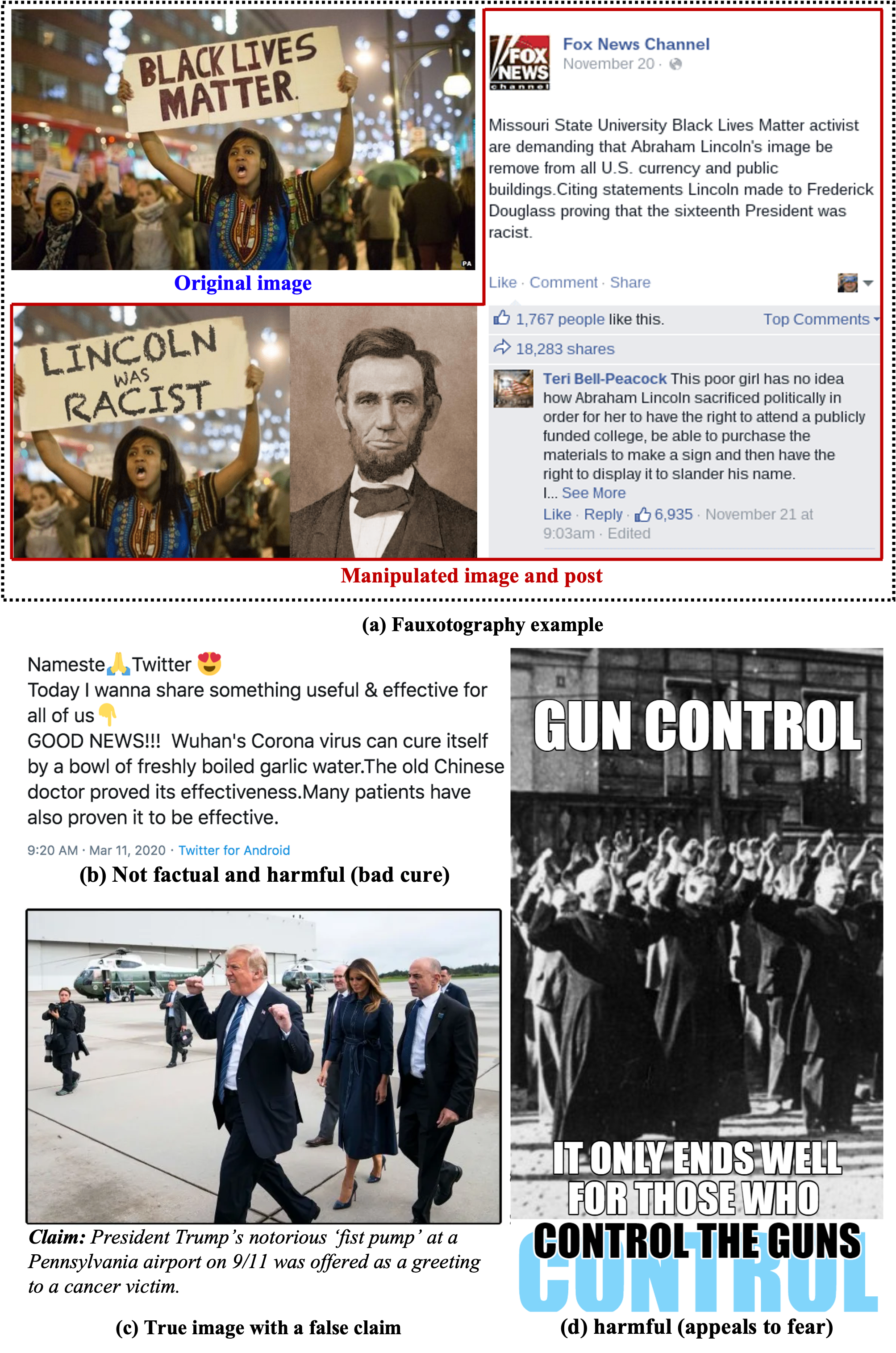}
\caption[fauxo]{Examples of textual and visual contents that show (a) \textit{fauxotographic content} (which is both harmful and false),
\footnote{\url{https://www.snopes.com/fact-check/abe-lincoln-racist-protest-sign/}} 
(b) harmful content promoting bad cure (text-only, and false),  (c) true image with a false claim about it (malicious), and (d) harmful content, where the text and the image collectively appeal to fear.}
\label{fig:example_fact_harmful}
\end{figure}

In Figure \ref{fig:example_fact_harmful}, we provide examples textual and visual content that are harmful and false, true image with false claim, and harmful meme. 

\footnotetext{\url{https://www.snopes.com/fact-check/abe-lincoln-racist-protest-sign/}}

\begin{figure}[]
\centering
\includegraphics[width=0.24\textwidth]{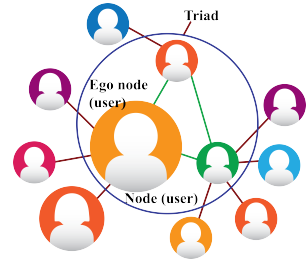}
\caption{Example of social network with users. \textbf{Node:} A node can be a users or a spreader. \textbf{Ego:} ``Ego'' is an individual ``focal'' node (central user) and the nodes that are directly connected to it are called ``alters/spreaders.''
\textbf{Triad:} It (a set of three connected users) is the most basic subgraph of the network. \textbf{Community:} A community structure refers to the occurrence of groups of nodes in a network that are more densely connected internally than with the rest of the network.
}
\label{fig:example_social_network}
\end{figure}

\section{Modeling Techniques}
\label{sec:app_modeling}
\begin{figure}[h]
\includegraphics[width=0.9\columnwidth]{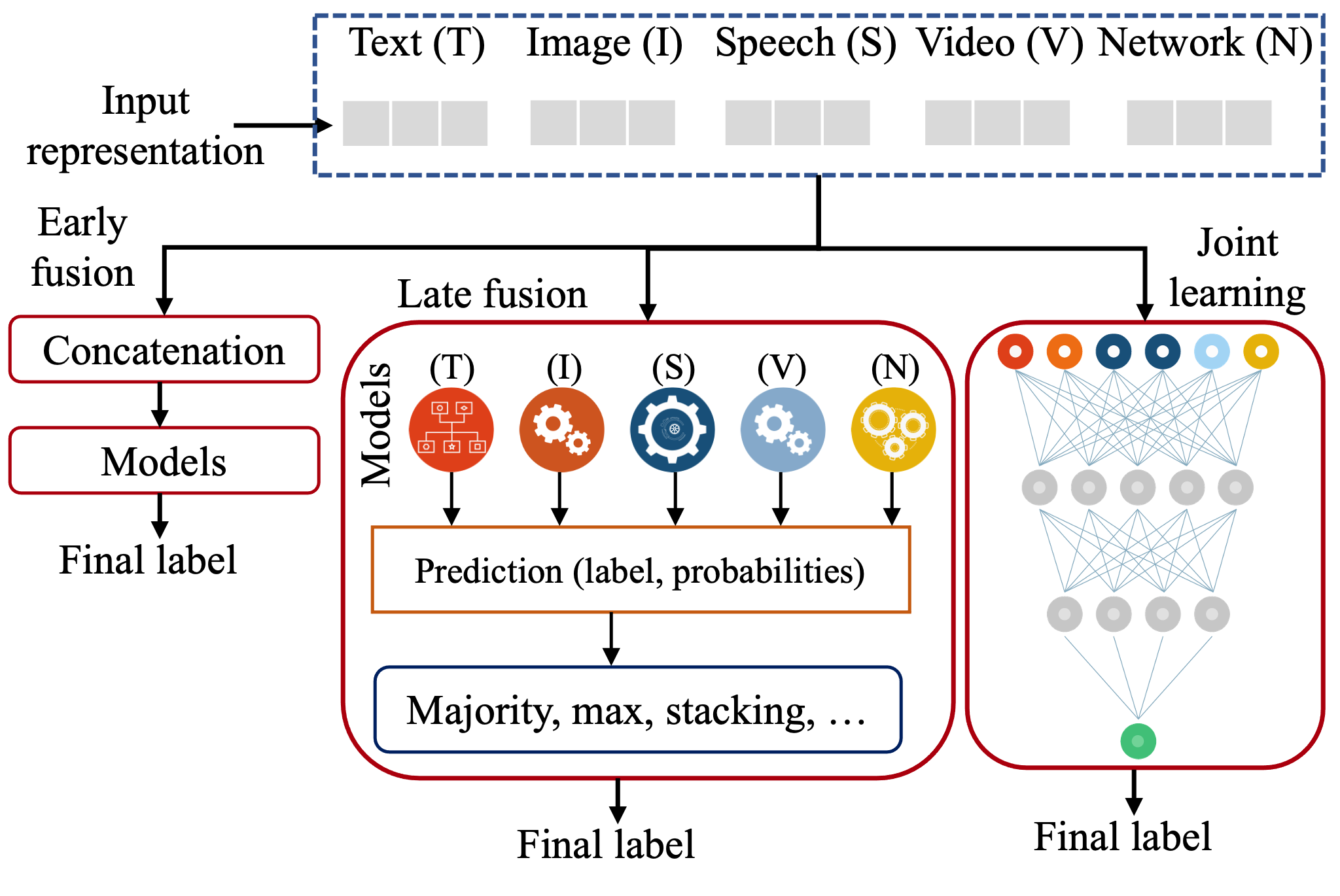}
\centering
\caption{
Multimodal approaches, including early and late fusion, and joint modal learning. The hybrid approach (combining early and late fusion) is not shown.
}
\label{fig:multimodal_arch}
\end{figure}

Figure \ref{fig:multimodal_arch} shows various multimodal approaches that have been proposed in the literature.

\section{Lessons Learned}
\label{sec:lessions}

\begin{enumerate} 
\item A lot of progress has been made on the problem, but the two components in the definition of disinformation (falseness and harmfulness) have been considered mostly in isolation. We argue that there is a need for tight integration of the factuality and the intentional harmfulness into the same detection model. 
These two aspects have been addressed together in \cite{alam2020fighting}, which shows that 56\% of Arabic false content is also harmful.
From Tables \ref{tab:related_studies_factuality} and \ref{tab:related_studies_harmfulness}, 
we observe that most multimodal datasets cover just 2--3 modalities, which combine some approaches depicted in Figure~\ref{fig:multimodal_arch}. Moreover, no multimodal dataset looks at both aspects of disinformation: factuality and harmfulness. While \citet{alam2020fighting} did address both aspects, they only covered the text modality. 
\item In the early phase of (dis)information spreading, user and content features are those that provide the highest contribution for detecting factuality. Indeed, at that time, a few interactions with content are available and the propagation network is small and sparse. As information spreads, the contribution of content-derived features remains constant, while propagation-derived features become richer and more informative. In summary, early prediction of factuality and veracity must necessarily rely heavily on users and content -- be it text, image, audio or video. Instead, analyses carried out at later times benefit more from network 
and temporal data. 
In the past decade, research on multimodality has shown its potential in several fields, which include audio-visual fusion \cite{mroueh2015deep,zhu2021deep,10.1088/1742-6596/1237/2/022144}, emotion recognition \cite{chen2021heu}, image and video captioning \cite{liu2021cptr}, multimedia retrieval and visual question answering \cite{summaira2021recent}. For factuality, \citet{baly2020written} showed that combining different modalities such as text, speech, and metadata yields improved performance compared to using individual modalities. Similar phenomena have been observed for other tasks such as hateful memes \cite{kiela2020hateful}, and propaganda detection \cite{SemEval2021-6-Dimitrov}. 

\end{enumerate}

\section{More Challenges}
\label{sec:app_more_major_challenges}

\begin{enumerate}
\item \textbf{Contextualization.} Existing methods of disinformation detection are mostly non-contextualized, i.e.,~the broader context of a news article in terms of the responses of the readers and how the users perceive them are not captured. We argue that the response thread under a news, the underlying social network among users, the propagation dynamics of the news and its mentions across social media need suitable integration to better capture the overall perspective on the news. 
    
\item \textbf{Meta Information.} Along with the news and the context, other information such as the authenticity of the news, the credibility of the authors of the news, the factuality of the news also play an important role for disinformation detection. Moreover, detecting whether the disinformation attack is a coordinated effort or an individual activity would also help understanding its severity.

\item \textbf{Bias, Region, and Cultural Awareness.} The performance of most of the existing systems is limited to the underlying dataset, particularly to the demography and the underlying cultural aspects. For instance, a model trained on an Indian political dataset may not generalize well to a US health-related dataset \cite{fortuna2021well}. 

\item \textbf{Disinformation on Evolving Topics.} Often, claims or harmful content are disseminated based on the current event; information about COVID-19 and vaccines are examples of such use cases. Existing models might fail on such use cases, and thus zero-shot or few-shot learning might be an important future avenue to explore.

\item \textbf{Transparent and Accountable Models.} 
The detection models should be designed in a way that their outcomes are unbiased and more accountable to ethical considerations. The models for disinformation detection should present the outcome in such a way that a practitioner can interpret it and understand why a piece of information is flagged as disinformation, what is the related real news based on which the judgment was made, and which part of the information was counterfeit. There is also a lack of datasets containing disinformation with explanations and the corresponding real information. 

\item \textbf{Fine-grained Detection.} Existing disinformation detection models are mostly binary classifiers: given a piece of news, they aim to detect whether it is a disinformation or not. Such binary signals might be enough in certain cases. However, in many other cases, more fine-grained labels can help to make a better decision. 
For example, whether a social media post is fake or genuine can help fact-checkers, but having more fine-grained information such as true, satire/parody, misleading, manipulated, false connection, or imposter content can be even more helpful \cite{nakamura2019r}.
Therefore, rather than a binary classification, one could cast the problem as a multi-class classification task or even an ordinal regression, or just a regression task. This would also help prioritize disinformation for reactive measurements.
\end{enumerate}

\end{document}